\documentclass[12pt]{article}                     
\usepackage{graphicx}

\usepackage{amsbsy,amsfonts}

\newcommand{\be}{\begin{equation}}
\newcommand{\ee}{\end{equation}}
\newcommand{\ba}{\begin{array}}
\newcommand{\ea}{\end{array}}
\newcommand{\p}{\partial}

\newcommand{\ds}{\displaystyle}
\newtheorem{lem}{Lemma}

\newtheorem{theo}{Theorem}
\newfont{\MSBM}{msbm10}
\newfont{\msbm}{msbm7}
\date{}

\begin{document}

\title{\protect\vspace{-15mm}\Large\bf Infinitely many local higher symmetries\\ without recursion operator or master symmetry:\\
integrability of the Foursov--Burgers system revisited}



\vspace{-5mm}
\author{{\sc Artur Sergyeyev}\\
              Mathematical Institute, Silesian University in Opava \\
              Na Rybn\'\i{}\v{c}ku 1, 74601 Opava, Czech Republic\\
              E-mail {\tt Artur.Sergyeyev@math.slu.cz}           
}

\date{January 19, 2008}

\maketitle
\vspace{-17mm}
\begin{abstract}
We consider the Burgers-type system studied by Foursov,
\[
\begin{array}{lcl}
w_t &=& w_{xx} + 8 w w_x + (2-4\alpha )z z_x,\vspace{3mm}\\
z_t &=& (1-2\alpha )z_{xx} - 4\alpha z w_x + (4-8\alpha )w z_x -
(4+8\alpha )w^2 z + (-2+4\alpha )z^3,
\end{array} 
\]
for which no recursion operator or master symmetry was known so far,
and prove that this system admits infinitely many local
higher symmetries that are constructed using a nonlocal {\em
two-term} recursion relation rather than a recursion operator.


\end{abstract}

\section*{Introduction}
Following Foursov~\cite{Four} consider a Burgers-type system
\begin{equation}\label{fs}
\begin{array}{lcl}
w_t &=& w_{xx} + 8ww_x + (2-4\alpha )zz_x,\vspace{3mm}\\
z_t &=& (1-2\alpha )z_{xx} - 4\alpha zw_x + (4-8\alpha )wz_x -
(4+8\alpha )w^2z + (-2+4\alpha )z^3,
\end{array}
\end{equation}
where $\alpha$ is a real parameter.

For $\alpha=0$ system (\ref{fs}) is equivalent \cite{Four} to a
system found by Svinolupov~\cite{Svi} while for $\alpha = 1$ this
system is equivalent to system (4.13) in Olver and
Sokolov~\cite{OaS}. Finally, for $\alpha=1/2$  a recursion operator
for (\ref{fs}) was found in \cite{Four}.

Moreover, for any value of $\alpha$
the differential substitution
\begin{equation}\label{ds}
\begin{array}{rl}
w= \displaystyle\frac{u_{x}}{4u},\qquad z=
\displaystyle-\frac{v}{2\sqrt{u}},
\end{array}
\end{equation}
the inverse of which was found in \cite{tw},
maps the triangular system
\begin{equation}\label{ts}
u_t = u_{xx} +(1-2\alpha )v^2, \qquad 
v_t = (1-2\alpha )v_{xx}
\end{equation}
into (\ref{fs}).
\looseness=-1

System (\ref{ts}) is triangular and hence $C$-integrable.
$C$-integrability means here that solving (\ref{ts}) boils down to
solving a {\em linear} heat equation for $v$, $v_t=(1-2\alpha
)v_{xx}$, and then solving a {\em linear} inhomogeneous PDE, namely
$u_t=u_{xx} +(1-2\alpha )v^2$ for a given $v$. More generally,
$C$-integrability of a system of PDEs is tantamount to existence of
a transformation that reduces this system to a linear (or at least
triangular) one, see \cite{Cal} for further details.\looseness=-1

In view of the above system (\ref{fs}) is $C$-integrab\-le as well, and the general
solution of (\ref{fs}) is obtained by plugging the general solution
of (\ref{ts}) into (\ref{ds}).
However, we are left with an open problem of whether the system
(\ref{fs}) has infinitely many local higher symmetries for
arbitrary value of $\alpha$. Recall that geometrically
a higher (or generalized \cite{olv_eng2})
symmetry for a system of PDEs is essentially a solution of
linearized version of this system; the solution in question is a vector function
on the associated diffiety and, in general, depends on higher (i.e., of order greater than one)
jets, see e.g.\ \cite{krasbook}, \cite{ibr}, \cite{vkl}, \cite{kk}, \cite{olv_eng2}
and references therein for further details. As for generalizations of this concept,
see \cite{krasbook}, \cite{ibr}, \cite{kon_mok}, \cite{vk}, \cite{vin1}, \cite{vkl}, \cite{kk}
and references therein for nonlocal higher symmetries,
and \cite{fl}, \cite{kl}, \cite{kl2}, \cite{bk},  \cite{rz}
and references therein for higher conditional symmetries.\looseness=-1

In a somewhat more technical language, a (local) higher symmetry for
(\ref{fs}) can be identified (see e.g.\ \cite{krasbook}, \cite{dor},
\cite{ibr}, \cite{vkl}, \cite{mik}, \cite{olv_eng2}) with a
two-component vector $\boldsymbol{K}=(K^1,K^2)^T$ whose entries
$K^i$ depend on $x,t,w,z$ and on a finite number of $x$-derivatives
of $w$ and $z$ up to an order $k$ (which is in general different for
different symmetries); here and below the superscript $T$ indicates
the transposed matrix. It is further required that the evolution
system
\be\label{symsys} w_{\tau}=K^1,\qquad z_{\tau}=K^2 \ee
is a commuting flow for (\ref{fs}), i.e.,
\[
\ds\frac{\p^2 w}{\p\tau\p t}=\frac{\p^2 w}{\p t\p\tau},\quad
\ds\frac{\p^2 z}{\p\tau\p t}=\frac{\p^2 z}{\p t\p\tau},
\]
where the partial derivatives with  respect to $t$ and $\tau$ are computed
using (\ref{fs}) and (\ref{symsys}).

For $\alpha\neq 0,1/2,1$
system (\ref{fs}) was known to have \cite{Four} six higher
symmetries but no recursion operator or master symmetry
was found so far, so it was not clear whether
(\ref{fs}) for $\alpha\neq 0,1/2,1$ has infinitely many
(rather than just six) higher symmetries.
Recall that a recursion operator is, in essence, an operator that maps
symmetries into (new) symmetries, see e.g.\
\cite{bl}, \cite{krasbook}, \cite{dor}, \cite{ibr}, \cite{kk}, \cite{olv_eng2}
and references therein. In turn,
a master symmetry is essentially a higher symmetry, typically a {\em nonlocal} one,
such that the commutator of this symmetry with any given symmetry
yields a (new) symmetry, see e.g.\ \cite{bl}, \cite{dor}, \cite{ff}, \cite{krasbook},
\cite{oevth}, \cite{olv_eng2} and references therein for details.

In view of the recent results of Sanders and van der Kamp \cite{svk}
who found several examples of two-component triangular evolution
systems that possess only a finite number (greater than one) of
local higher symmetries, it is natural to ask whether
(\ref{fs}) could provide an example of a {\em non-triangular}
system with finitely many local higher symmetries.\looseness=-1

In the present paper we show that this is not the case: system
(\ref{fs}) has infinitely many commuting local higher
symmetries. Quite unusual, however, is the fact that these
symmetries are generated using a {\em nonlocal two-term recursion
relation} (\ref{rec}) rather than a recursion operator or a master
symmetry, see Theorem~\ref{srt} below for details. Moreover, we
believe that it is impossible to construct a recursion operator of a
reasonably ``standard" form that would reproduce at least a part of the
hierarchy from Theorem~\ref{srt}, but no proof of this claim is
available so far.\looseness=-1

Generation of symmetries via recursion relations turns out to be of
considerable interest on its own right. To the best of our
knowledge, the first examples of this kind have appeared in
\cite{BSW}, with the simplest case given by Eqs.(\ref{symtr}),
(\ref{symtrec0}) below. However, the recursion relation
(\ref{symtr}) is {\em local}, and therefore obviously produces local
symmetries; for other recursion relations in \cite{BSW} locality of
symmetries is also pretty much immediate, unlike the recursion
relation (\ref{rec}) in Theorem~\ref{srt} below.\looseness=-1

In general, if we deal with a recursion relation that involves
nonlocalities, then establishing locality and commutativity of the
symmetries generated using this relation is a highly nontrivial
task. It is important to stress that the hitherto known methods for
proving locality of hierarchies of symmetries are based upon
existence of a hereditary recursion operator or of a master
symmetry, see e.g.\ \cite{sw}, \cite{serg02}, \cite{serg05},
\cite{wan08} and references therein, in general are not applicable
in this situation.\looseness=-1

For the particular case of recursion relation (\ref{rec}) we proved locality
and commutativity of the symmetries $\boldsymbol{K}_n$ (\ref{rec}) using some
{\em ad hoc} arguments in spirit of \cite{ibr} and \cite{krasprep},
but it would be very
interesting to find general, more powerful methods that would not
require the existence of scaling symmetry.\looseness=-1

It would be of interest to understand geometrical meaning of
commutativity of symmetries in this setting. For the ``standard"
hierarchies this means vanishing of the Nijenhuis torsion of the
recursion operator (see e.g.\ \cite{dor}) but it is not quite
clear how one could generalize this to the case of the recursion
relations.\looseness=-1

Finally, perhaps the most important open problem here is whether there exist
any $S$-integrable (i.e., roughly speaking, integrable via
the inverse scattering transform, see \cite{Cal} for details)
systems of PDEs that have no ``standard" recursion operator or master symmetry
but nevertheless possess infinitely many symmetries generated
through a recursion relation involving two or more terms.

\section{Preliminaries}
We start with recalling some basic properties of (\ref{ts}).

First of all, for $\alpha=1/2$ this system decouples and takes the form
\begin{equation}\label{tso}
u_t = u_{xx},\qquad v_t = 0,
\end{equation}
i.e., in this case
we have a linear homogeneous heat equation for
$u$, and $v$ is simply an arbitrary function of $x$.\looseness=-1


On the other hand, for $\alpha\neq 1/2$ upon introducing a new
independent variable $\tau=(1-2\alpha) t$ instead of $t$ in
(\ref{ts}) and setting $a=1/(1-2\alpha)$ we obtain the system that
was studied, {\em inter alia}, in \cite{BSW}, viz.
\begin{equation}\label{ts1}
u_\tau = a u_{xx} + v^2,\qquad v_\tau = v_{xx}.
\end{equation}

Moreover, system (\ref{ts1}) (and hence (\ref{ts})) has infinitely
many symmetries. Indeed, by Theorem 2.2 of \cite{BSW} the system in
question has infinitely many local higher symmetries of the form
\begin{equation}\label{symtr}
\boldsymbol{G}_n = \left(\begin{array}{c}
b_n u_n + Q_n\\
v_n \end{array}\right),
\end{equation}
where $b_n = b_{n-1} - (1-a)b_{n-2}/2$ with $b_1 = 1$, $b_2 = a$,
and
\begin{equation}\label{symtrec0}
Q_n = D_x Q_{n-1}-\frac{1-a}{2}D_{x}^2Q_{n-2} + v v_{n-2}.
\end{equation}
The initial conditions for the recursion (\ref{symtrec0}) are $Q_1 =
0$, $Q_2 = v^2$.

Here $u_i$ and $v_i$ stand for the $i^{\mathrm{th}}$ $x$-derivatives
of $u$ and $v$, $u_0\equiv u$, $v_0\equiv v$, and $D_x$ denotes the
total $x$-derivative (see e.g.\ \cite{krasbook}, \cite{ibr}, \cite{mik}, \cite{olv_eng2})
\[
D_x=\displaystyle \frac{\partial}{\partial
x}+\sum\limits_{i=0}^\infty \left(u_{i+1}\displaystyle
\frac{\partial}{\partial u_i}+ v_{i+1}\displaystyle
\frac{\partial}{\partial v_i}\right).
\]
Recall that a function $f$ of $x,t,u_0,v_0,u_1,v_1,\dots$ is said to
be {\em local} if it depends only on $x$, $t$, and a {\em finite}
number of $u_i$ and $v_i$, see e.g.\ \cite{krasbook}, \cite{mik},
\cite{olv_eng2} and references therein for further details.
\looseness=-1


Note that for $a=1$ (i.e., $\alpha=0$) a recursion operator for
(\ref{ts1}) was found by Oevel in \cite{oevth}, so for $\alpha=0$ we
can readily find a recursion operator for (\ref{fs}) from the
Oevel's recursion operator. On the other hand, for $\alpha=1/2$ (i.e., $a=0$) Foursov
\cite{Four} also found a recursion operator for (\ref{fs}).

However, for
generic $a$ no recursion operator is known for (\ref{ts1}).
Moreover, for generic $a$ the form of leading coefficients of higher-order
symmetries of (\ref{ts1}), see (\ref{symtr}),
appears to preclude existence of any ``reasonable"
recursion operator for (\ref{ts1}), and
thus for (\ref{ts}) and (\ref{fs}) as well, but so far
we were unable to prove this claim.
The supposed nonexistence of recursion operator should be closely related to the
number-theoretic aspects of the symmetry analysis for (\ref{ts1}),
see \cite{BSW}, \cite{sw08} and references therein. \looseness=-1

\section{Infinitely many local higher symmetries for (\ref{fs})}
Using the inverse of the differential substitution (\ref{ds}) (of course,
the inverse in question
should be considered as a covering in the sense of
\cite{krasbook}, \cite{vk}, \cite{vkl}, \cite{vin1}, \cite{kk} and references therein)
we can obtain infinitely many symmetries for (\ref{fs}) with an
arbitrary value of $\alpha$ different from $1/2$
from the symmetries of (\ref{ts1}) given
by (\ref{symtr}), (\ref{symtrec0}).
Most importantly, all of these symmetries are local by virtue of the
following result.\looseness=-1
\begin{theo}\label{srt}
For $\alpha\neq 1/2$ system (\ref{fs}) has infinitely many
commuting {\em local} higher symmetries $\boldsymbol{K}_{n}$
generated using the nonlocal {\em two-term} recursion relation
\begin{equation}\label{rec}
\boldsymbol{K}_{n} =\left(\begin{array}{cc}
  D_{x} + 4 w + 4w_x D_{x}^{-1} & 0\vspace{2mm}\\
  2(z_x - 2 wz) D_{x}^{-1} &   D_{x}  + 2w\end{array}\right)
\boldsymbol{K}_{n-1} + M \boldsymbol{K}_{n-2}
\end{equation}
starting from 
\begin{equation}\label{ind}
\boldsymbol{K}_{1} = \left (\begin{array}{c}
w_{x}\\
z_{x}
\end{array}\right),
\qquad \boldsymbol{K}_{2} = \left (\begin{array}{c}
-\ds\frac{w_{xx} + 8 w w_x}{2 \alpha - 1} + 2z z_x\vspace{2mm}\\
z_{xx} + 4w z_x-\ds\frac{4z(\alpha w_x + (2 \alpha +1)w^2)}{2 \alpha
- 1} - 2z^3
\end{array}\right).
\end{equation}
Here
\[
M= \left(\begin{array}{lc}
M_{11}\quad &   z D_{x} + z_x\\[1ex]
M_{21}\quad &  - 2 z^2
\end{array}
\right),
\]
\[
\begin{array}{ll}
M_{11}=&-\ds\frac{1}{2\alpha-1} \left(\alpha D_{x}^{2}
 + 8 \alpha w D_{x}
  + 2(\alpha(6 w_x + 8 w^2) - (2\alpha-1)z^2) \right. \\[1.5ex]
  & \left. {}+ 4\left(\alpha(w_{xx} + 8 w w_x) - (2\alpha-1)z z_x\right)
 D_{x}^{-1}\right),\\[2ex]
M_{21}=& \ds\frac{2\alpha }{2 \alpha - 1}z D_{x} +
\ds\frac{16 \alpha}{2 \alpha - 1}w z +4 \left(\ds\frac{2 \alpha }{2
\alpha - 1}(w_x + 4 w^2) z - z^3\right) D_{x}^{-1}.
\end{array}
\]
\end{theo}
We defer the proof of this theorem until the next section.

Recall (see Introduction) that locality of the symmetries $\boldsymbol{K}_{j}$, $j\in\mathbb{N}$,
means that they depend only on $x,t,w,z$ and a {\em finite} number of their $x$-derivatives
$w_i$ and $z_i$ ($w_i$ and $z_i$ stand for the $i^{\mathrm{th}}$ $x$-derivatives
of $w$ and $z$, and we set $w_0\equiv w$ and $z_0\equiv z$ for convenience)
and do not involve any nonlocal quantities, cf.\ e.g.\ \cite{krasbook}, \cite{vkl}, \cite{kk},
\cite{mik}, \cite{olv_eng2}.\looseness=-1

The total $x$-derivative $D_x$ now takes the form
\[
D_x=\displaystyle \frac{\partial}{\partial
x}+\sum\limits_{i=0}^\infty \left(w_{i+1}\displaystyle
\frac{\partial}{\partial w_i}+ z_{i+1}\displaystyle
\frac{\partial}{\partial z_i}\right).
\]

Let $f\in\mathrm{Im} D_x$ be a polynomial in a finite number
of $w_i$ and $z_i$ with zero free term (let us stress that $f$ is not allowed
to depend explicitly on $x$ and $t$).
In Theorem~\ref{srt} and
below we make a {\em blanket assumption} that the result of action
of $D_x^{-1}$
on any such $f$ 
again is a polynomial in a finite number of $w_i$ and $z_i$ with
zero free term, i.e., we always set the integration constant to
zero. Note that an alternative (and more general) approach is to define the operator $D_x^{-1}$
in the fashion described in \cite{gut}, \cite{mar}, \cite{ser00}.

\looseness=-1

With the above assumption in mind we readily find that the
first nontrivial symmetry generated via (\ref{rec}), namely,
$\boldsymbol{K}_3$, has the form $\boldsymbol{K}_3=(K_3^1,K_3^2)^T$,
where
\[
\begin{array}{ll}\ds
K_3^1=&-\ds\frac{(\alpha + 1)}{2 \alpha - 1} w_{xxx} - \frac{12
(\alpha + 1) }{2 \alpha - 1}w w_{xx} + 3 z z_{xx} - \frac{12(\alpha
+ 1)}{2 \alpha - 1}w_x^2 \\[2ex]
&\ds {}+ 3 z_x^2 - 6 z^2 w_x- \frac{48 (\alpha+1)}{2 \alpha - 1}w^2 w_x + 12 w z z_x,\\[1.5ex]
\ds K_3^2=& z_{xxx} + \ds\frac{6\alpha}{2 \alpha - 1}z w_{xx} + 6 w
z_{xx} + 6 w_x z_x +\ds\frac{12(4 \alpha + 1)}{2 \alpha - 1}w z w_x + 6
(2 w^2 -
z^2) z_x\\[2ex]
& \ds {} + \frac{24(2\alpha+1)}{2 \alpha - 1} w^3 z -12 w z^3.
\end{array} 
\]
We see that the symmetries $\boldsymbol{K}_{j}$, $j=1,2,3$ are independent of $x$ and $t$.
Moreover, it is easily seen
that by virtue of the above definition of action
of $D_x^{-1}$ the symmetries $\boldsymbol{K}_{j}$
in fact do not depend explicitly on $x$ and $t$ for all $j\in\mathbb{N}$.

\section{Proof of Theorem \ref{srt}}
The recursion relation (\ref{symtrec0}) can be rewritten in the terms of
symmetries $\boldsymbol{G}_i$ from (\ref{symtr}) as
\begin{equation}\label{symtrec}
\boldsymbol{G}_n = D_x (\boldsymbol{G}_{n-1})-\left(\ba{cc}
(1-a)/2 & 0\\0 & 0 \ea \right)D_{x}^2 (\boldsymbol{G}_{n-2}) +
\left(\ba{cc} 0 & v\\0 & 0 \ea \right) \boldsymbol{G}_{n-2}, 
\end{equation}
and the recursion relation (\ref{rec}) readily follows from
(\ref{symtrec}) and (\ref{ds}).

Now we have to show that the symmetries $\boldsymbol{K}_j$ generated
via the recursion relation (\ref{rec}) with the initial data (\ref{ind}) are
{\em local} in the sense of the preceding section for all $j=3,4,\dots$.
This obviously
boils down to proving that they do not involve nonlocalities like
$y=D_x^{-1}(w)$.\looseness=-1

To this end we shall use induction on $n$ starting from $n=1$. In
view of the specific form of the recursion relation (\ref{rec}) we
only have to show that once $\boldsymbol{K}_{j-1}=(K_{j-1}^1,
K_{j-1}^2)^T$ and $\boldsymbol{K}_{j-2}=(K_{j-2}^1, K_{j-2}^2)^T$
are such that $K_{j-1}^1,K_{j-2}^1\in\mathrm{Im} D_x$ (and hence
$\boldsymbol{K}_{j}$ is local) then we have $K_{j}^1\in\mathrm{Im}
D_x$. This is obviously true for $j=3$, so we only need to establish
validity of the induction step. This is done using the following
result.\looseness=-1
\begin{lem}Fix $j\geq 3$ and assume that $\boldsymbol{K}_i$, $i=j-2,j-1$, are local, and
$K_{j-1}^1,K_{j-2}^1\in\mathrm{Im} D_x$.
Then $\boldsymbol{K}_j$ defined via (\ref{rec}) is local too, and we have
$K_{j}^1\in\mathrm{Im} D_x$.
\end{lem}

{\em Proof of the lemma.} As $K_{j-1}^1,K_{j-2}^1\in\mathrm{Im} D_x$ by assumption,
locality of $\boldsymbol{K}_j$ is immediately inferred from (\ref{rec}),
so it remains to prove that $K_{j}^1\in\mathrm{Im} D_x$.

Now, the condition $K_{j}^1\in\mathrm{Im} D_x$ can be restated as
follows: $\rho_j\in\mathrm{Im} D_x$, where
$\rho_j\equiv\rho_0'[\boldsymbol{K}_j]$, $\rho_0=w$ is a conserved
density for (\ref{fs}), and $f'[\boldsymbol{K}]$ stands (cf.\ e.g.\
\cite{bl}, \cite{dor}, \cite{oevth}) for the directional derivative
(also known as linearization, see e.g.\ \cite{krasbook}, \cite{vkl},
\cite{kk}) of $f$ along $\boldsymbol{K}=(K^1,K^2)^T$:
\[
f'[\boldsymbol{K}]=\sum\limits_{i=0}^\infty\left(\ds\frac{\partial
f}{\partial w_i}D_x^i(K^1)+ \ds\frac{\partial f}{\partial
z_i}D_x^i(K^2)\right).
\]
Recall (see e.g.\
\cite{krasbook}, \cite{ibr}, \cite{mik}, \cite{olv_eng2})
that a local function $\rho$ is said
to be a local {\em conserved density}
for (\ref{fs}) if $D_t(\rho)\in\mathrm{Im}
D_x$. A local conserved density $\rho$ is said to be {\em nontrivial} if
$\rho\not\in\mathrm{Im} D_x$, and {\em trivial} otherwise.\looseness=-1

Next, as $\boldsymbol{K}_j$ is a symmetry of (\ref{fs}) by
construction, the quantity $\rho_j$ is a local conserved density for
(\ref{fs}), see e.g.\ \cite{bl}, \cite{ibr}, \cite{krasbook},
\cite{olv_eng2}.

However, for $\alpha\neq 1, 1/2$ the quantity $\rho_0$ is the only
nontrivial local conserved density
for (\ref{fs}) modulo the terms from $\mathrm{Im} D_x$.
Indeed, by virtue of the results of \cite{mik} (see also
Theorem 5-1 of \cite{fol}, and cf.\ \cite{ps} and references therein) any nontrivial
local conserved density
for (\ref{fs}) depends (again modulo the terms from $\mathrm{Im} D_x$)
only on $x,t,w,z,w_x,z_x,\allowbreak
w_{xx},z_{xx}$, and the direct 
search for all conserved densities of this form proves our
claim.\looseness=-1

Hence the most general local conserved density for (\ref{fs}) has
the form $c \rho_0+\tilde\rho$, where $\tilde\rho\in\mathrm{Im} D_x$
and $c$ is a constant. In particular, if $\boldsymbol{K}_j$ is local
then we have
\begin{equation}\label{dd}
\rho_j=c_j \rho_0+\tilde\rho_j,
\end{equation}
where $\tilde\rho_j\in\mathrm{Im} D_x$ and $c_j$ are constants.

In analogy with Krasil'shchik \cite{krasprep},
let us show that if $\boldsymbol{K}_j$ is local then $c_j=0$,
and thus $\rho_j\in \mathrm{Im} D_x$.

First of all, note that system (\ref{fs}) has \cite{Four} a scaling
symmetry\footnote{Note that $\boldsymbol{S}$, $\boldsymbol{K}_1$,
and $\boldsymbol{K}_2$ exhaust all linearly independent Lie point
(or, more precisely, higher symmetries that are equivalent to the
Lie point ones in the sense of \cite{ibr,olv_eng2}) symmetries of
(\ref{fs}).}
\[
\boldsymbol{S} = 2t (1-2\alpha)\boldsymbol{K}_{2}+ x \boldsymbol{K}_{1}
+\left(\begin{array}{l}
w\\
z\end{array}\right),
\]
and it is readily seen that all $\boldsymbol{K}_{j}$ constructed
using (\ref{rec}) are $\boldsymbol{S}$-homogeneous: we have
$L_{\boldsymbol{S}}(\boldsymbol{K}_{j})\equiv\lbrack\boldsymbol{S},\boldsymbol{K}_{j}\rbrack
=j \boldsymbol{K}_{j}$, $j=1,2,\dots$. Here $L_{\boldsymbol{Q}}$
denotes the Lie derivative along ${\boldsymbol{Q}}$, see e.g.\
\cite{bl}, \cite{olv_eng2}, \cite{sw}, \cite{serg05} for details,
and $\lbrack\cdot,\cdot\rbrack$ is the standard commutator of
symmetries (see e.g.\ \cite{bl}, \cite{krasbook}, \cite{dor},
\cite{ibr}, \cite{kk}, \cite{mik}, \cite{olv_eng2}):\looseness=-1
\begin{equation}\label{com}
\lbrack\boldsymbol{F},\boldsymbol{G}\rbrack=
\boldsymbol{G}'[\boldsymbol{F}]-\boldsymbol{F}'[\boldsymbol{G}].
\end{equation}

Hence $L_{\boldsymbol{S}}(\rho_{j})=\rho_j'[\boldsymbol{S}]=(j+1)
\rho_{j}+\eta_j$, where $\eta_j\in\mathrm{Im} D_x$. On the other
hand, 
\[
L_{\boldsymbol{S}}(c_j \rho_{0}+\tilde\rho_j)=c_j\rho_0+\zeta_j,
\]
where $\zeta_j\in\mathrm{Im} D_x$.
Therefore, if we  act by $L_{\boldsymbol{S}}$ on the left- and
right-hand side of (\ref{dd}) and equate the resulting expressions,
we obtain
\begin{equation}\label{imd}
j c_j\rho_0+\theta_j=0,
\end{equation}
where $\theta_j\in\mathrm{Im} D_x$.

As $\rho_0\not\in\mathrm{Im} D_x$ and $j\neq 0$ by assumption,
Eq.(\ref{imd}) can hold only if $c_j\equiv 0$. Hence
$\rho_j=\tilde\rho_j\in\mathrm{Im} D_x$, and the lemma is proved.
$\square$

Now that we have proved locality of $\boldsymbol{K}_j$ for all
$j\in\mathbb{N}$, we can easily prove commutativity of
$\boldsymbol{K}_j$ with respect to the bracket (\ref{com}):
$\lbrack\boldsymbol{K}_i, \boldsymbol{K}_j\rbrack=0$,
$i,j=1,2,3,\dots$.

First of all,
in analogy with the reasoning presented in Chapter 4 of \cite{ibr}
for scalar evolution equations (cf.\ also
\cite{krasbook}, \cite{krasprep}, \cite{vkl}) it can be shown that any
$\boldsymbol{S}$-homogeneous
$x,t$-independent local higher symmetry $\boldsymbol{G}$ of
(\ref{fs}) of order $j\geq 1$
is of the form 
\begin{equation}\label{kj}
\boldsymbol{G}=\left(\ba{c}\alpha_j w_j+\tilde{G}_j^1\\[1ex]
\beta_j z_j+\tilde{G}_j^2\ea\right),
\end{equation}
where $\alpha_j,\beta_j$ are constants, and $\tilde{G}_j^1,
\tilde{G}_j^2$ are polynomials in
$w, z,  w_1, z_1, \dots,\allowbreak w_{j-1}, z_{j-1}$, and 
these polynomials have no free terms and no linear terms.

Let $\mathcal{L}_k$ be the space of $\boldsymbol{S}$-homogeneous
$x,t$-independent local higher symmetries of (\ref{fs}) of
order no greater than $k$. By the above, there exists a basis in
$\mathcal{L}_k$ that consists of symmetries of the form (\ref{kj})
for $j=1,2,\dots,k$ (note that in principle this basis may contain
more than one symmetry of given order $j$).

Clearly, $\boldsymbol{K}_i\in\mathcal{L}_i$ and
$\lbrack\boldsymbol{K}_i,
\boldsymbol{K}_j\rbrack\in\mathcal{L}_{i+j}$. But using (\ref{com})
and (\ref{kj}) we readily see that the commutator
$\lbrack\boldsymbol{K}_i, \boldsymbol{K}_j\rbrack$ contains no
linear terms, and hence (cf.\ e.g.\ \cite{ibr}) this commutator can
belong to $\mathcal{L}_{i+j}$ only if $\lbrack\boldsymbol{K}_i,
\boldsymbol{K}_j\rbrack=0$. Thus, $\lbrack\boldsymbol{K}_i,
\boldsymbol{K}_j\rbrack=0$ for all $i,j\in\mathbb{N}$, and this
completes the proof of Theorem~\ref{srt}. 
\looseness=-1

\subsection*{Acknowledgements}
This research was supported in part by the Czech Grant Agency
(GA\v{C}R) under grant No.\ 201/04/0538, by the Ministry of
Education, Youth and Sports of the Czech Republic (M\v SMT \v CR)
under grant MSM 4781305904, and by Silesian University in Opava
under grant IGS 9/2008. It is my great pleasure to thank B. Kruglikov,
M. Marvan and T. Tsuchida for useful suggestions.
\looseness=-1

{

}

\begin{thebibliography}{99}
\bibitem{BSW} F. Beukers, J.A. Sanders, J.P. Wang, On
Integrability of Systems of Evolution Equations, {\em J. Diff. Eq.}
\textbf{172} (2001), 396--408.

\bibitem{bl} M. B\l aszak, {\em Multi-Hamiltonian Theory of Dynamical
Systems}, Springer, Heidelberg, 1998.\looseness=-1

\bibitem{krasbook}A.V. Bocharov et al.,
{\em Symmetries and conservation laws for differential equations of mathematical physics}.
Edited and with a preface by I.S. Krasil'shchik and A.M. Vino\-gradov,
American Mathematical Society, Providence, RI, 1999.

\bibitem{Cal}
F. Calogero, Why are certain nonlinear PDEs both widely applicable and integrable?,
in: {\em What is
Integrability?}, ed. V.E.~Zakharov, Springer, New York, 1991,
pp.~1--62.\looseness=-1

\bibitem{dor}I. Dorfman,
{\it Dirac Structures and Integrability of Nonlinear Evolution
Equations},
John Wiley \& Sons, Chichester, 1993.


\bibitem{ff} A.S. Fokas and B. Fuchssteiner,
The hierarchy of the Benjamin--Ono equation,
{\it Phys. Lett. A} {\bf 86} (1981), 341--345.

\bibitem{fl}A.S. Fokas, Q.M. Liu,
Generalized conditional symmetries and exact solutions of non-integrable equations.
{\em Theor. Math. Phys.} {\bf 99} (1994), no. 2, 571--582.

\bibitem{fol}K. Foltinek,
Conservation Laws of Evolution Equations: Generic Nonexistence, {\em
J. Math. Anal. Appl.} {\bf 235} (1999), 356--379.

\bibitem{Four} M.V. Foursov, On integrable coupled Burgers-type
equations, {\it Physics Letters A} {\bf 272} (2000) 57--64.

\bibitem{gut}G.A. Guthrie, Recursion operators and non-local symmetries,
{\it Proc. Roy. Soc. London Ser. A} {\bf 446} (1994), No.1926, 107--114.

\bibitem{ibr} N.H. Ibragimov, {\em Transformation Groups Applied to
Mathematical Physics}, Reidel, Dordrecht 1985.\looseness=-1


\bibitem{svk}P.H. van der Kamp, J.A. Sanders,
Almost integrable evolution equations, {\em Selecta Math. (N.S.)}
{\bf 8} (2002), no. 4, 705--719.


\bibitem{kon_mok}B.G. Konopelchenko, V.G. Mokhnachev,
On the group-theoretical analysis of differential equations. 
{\em Soviet J. Nuclear Phys.} {\bf 30} (1979), no. 2, 559--567.

\bibitem{vk}I.S. Krasil'shchik, A.M. Vinogradov,
Nonlocal symmetries and the theory of coverings: an addendum to
A. M. Vinogradov's Local symmetries and conservation laws,
{\em Acta Appl. Math.} {\bf 3} (1984), 79--96.

\bibitem{vkl}I.S. Krasil'shchik, V.V. Lychagin, A.M. Vinogradov,
{\em Geometry of jet spaces and nonlinear partial differential equations},
Gordon and Breach, New York, 1986.

\bibitem{vin1}I.S. Krasil'shchik, A.M. Vinogradov, Nonlocal trends
in the geometry of differential equations: symmetries, conservation laws,
and B\"acklund transformations. Symmetries of partial differential equations, Part I.
{\em Acta Appl. Math.} {\bf 15} (1989), no. 1-2, 161--209.

\bibitem{kk}I.S. Krasil'shchik, P.H.M. Kersten,
{\em Symmetries and recursion operators for classical and supersymmetric differential equations.}
Kluwer Academic Publishers, Dordrecht, 2000. 

\bibitem{krasprep}I. Krasil'shchik,
A simple method to prove locality of symmetry hierarchies,
Preprint DIPS 9/2002, available online at {\tt http://www.diffiety.org/}

\bibitem{kl}B. Kruglikov, V. Lychagin, Mayer brackets and solvability of PDEs II,
   {\em Trans. A.M.S.} {\bf 358} (2006), no. 3, 1077--1103.

\bibitem{kl2}B. Kruglikov, V. Lychagin, Compatibility, multi-brackets and integrability
   of systems of PDEs, arXiv: math/0610930

\bibitem{bk}B. Kruglikov, Symmetry approaches for reductions of PDEs,
   differential constraints and Lagrange-Charpit method,
   {\em Acta Appl. Math.} {\bf 101} (2008), 
   145--161 (arXiv: 0712.3425).

\bibitem{mar}M. Marvan, Another look on recursion operators, in:
{\em Differential Geometry and Applications (Brno, 1995)}, eds. J.
Jany\v ska et al.,
Masaryk University, Brno, 1996,
pp.~393--402, available online at {\tt
http://www.emis.de/proceedings}



\bibitem{mik}
A.V. Mikhailov, A.B. Shabat and V.V. Sokolov, The symmetry approach
to classification of integrable equations, in: {\em What is
Integrability?}, ed. V.E.~Zakharov, Springer, New York, 1991,
pp.~115--184.\looseness=-1


\bibitem{oevth}
W. Oevel, {\it Rekursionmechanismen f\"ur Symmetrien und
Erhaltungss\"atze in Integrablen Systemen}, Ph.D. thesis, University
of Paderborn, Paderborn, 1984. \looseness=-1

\bibitem{olv_eng2} P.J. Olver, {\em Applications of Lie Groups to Differential Equations},
Springer, New York,
1993.


\bibitem{OaS} P.J. Olver, V.V. Sokolov, Integrable evolution equations on associative algebras,
{\em Commun. Math. Phys.} {\bf 193} (1998) 245--268.


\bibitem{ps}R.O. Popovych and A.M. Samoilenko,  Local conservation laws of second-order evolution equations,
{\em J. Phys. A: Math. Theor.} {\bf 41} (2008), 362002, 11 p. (arXiv:0806.2765).


\bibitem{sw}
J.A. Sanders and J.P. Wang, Integrable Systems and their Recursion
Operators, {\em Nonlinear Analysis} {\bf 47} (2001), no.8,
5213--5240.

\bibitem{sw08}J.A. Sanders and J.P. Wang, Number Theory and the Symmetry
Classification of Integrable Systems, in: {\em Integrability}, ed. A.V. Mikhailov,
Springer, Berlin etc., 2009, p.89--118; draft version available online at
{\tt\verb|http://www.cs.vu.nl/~jansa/ftp/WORK100/Chapter2_SandersWang.pdf|}

\bibitem{ser00}A. Sergyeyev, On recursion operators
and nonlocal symmetries of evolution equations, in:
{\em Proc. Sem. Diff. Geom.}, ed. D. Krupka, Silesian
University in Opava, Opava, 2000, pp.~159--173 (arXiv: nlin.SI/0012011).

\bibitem{serg02}A. Sergyeyev,  On sufficient conditions of locality
for hierarchies of symmetries of evolution systems, {\em Rep. Math.
Phys.} {\bf 50} (2002), no.3, 307--314.


\bibitem{serg05}A. Sergyeyev, Why nonlocal recursion operators produce local symmetries:
new results and applications, {\em J. Phys. A: Math. Gen.} {\bf 38} (2005), 
3397--3407 (arXiv: nlin.SI/0410049).

\bibitem{Svi} S.I. Svinolupov, On the analogues of the Burgers equation,
{\em Phys. Lett. A} {\bf 135} (1989) 32--36.

\bibitem{tw}T. Tsuchida and T. Wolf, Classification of polynomial integrable systems of
mixed scalar and vector evolution equations: I, {\em J. Phys. A:
Math. Gen.} {\bf 38} (2005), 7691--7733 (arXiv: nlin.SI/0412003).

\bibitem{wan08}J.P. Wang, Lenard scheme for two-dimensional periodic Volterra chain,
 arXiv:0809.3899

\bibitem{rz}R.Z. Zhdanov,
Conditional Lie-B\"acklund symmetry and reduction of evolution equations,
{\em J. Phys. A} {\bf 28} (1995), no. 13, 3841--3850 (arXiv: solv-int/9505006).

\end{thebibliography}
\end{document}